\documentclass[twocolumn,showpacs,preprintnumbers,amsmath,amssymb,pre]{revtex4}

\usepackage{graphicx}
\usepackage{dcolumn}
\usepackage{bm}
\usepackage{enumerate}

\begin{document}

\title{Cattaneo--type subdiffusion--reaction equation}

\author{Tadeusz Koszto{\l}owicz}
 \email{tadeusz.kosztolowicz@ujk.edu.pl}
 \affiliation{Institute of Physics, Jan Kochanowski University,\\
         ul. \'Swi\c{e}tokrzyska 15, 25-406 Kielce, Poland.}

\date{\today}

\begin{abstract}
Subdiffusion in a system in which mobile particles $A$ can chemically react with static particles $B$ according to the rule $A+B\rightarrow B$ is considered within a persistent random walk model. This model, which assumes a correlation between successive steps of particles, provides hyperbolic Cattaneo normal diffusion or fractional subdiffusion equations for a system without chemical reactions. Starting with the difference equation, which describes a persistent random walk in a system with chemical reactions, using the generating function method and the continuous time random walk formalism, we will derive the Cattaneo--type subdiffusion differential equation with fractional time derivatives in which the chemical reactions mentioned above are taken into account. We will also find its solution over a long time limit. Based on the obtained results, we will find the Cattaneo--type subdiffusion--reaction equation in the case in which mobile particles of species $A$ and $B$ can chemically react according to a more complicated rule.
\end{abstract}

                            
\maketitle

\section{Introduction}

Subdiffusion--reaction equations have been studied extensively during the last decade \cite{zb,bah,yuste,kosztlew,seki,sung,s,sokolov,mendez,zoia,henry,yadav,vlad,eule}. Subdiffusion occurs in a medium where the mobility of particles is strongly
hindered due to the internal structure of the medium as, for
example, in porous media or gels \cite{mk,kdm}.
Subdiffusion can be treated as a random walk process which is characterized by the relation
\begin{equation}\label{eq1}
\left\langle (\Delta x)^2\right\rangle =\frac{2D_\alpha}{\Gamma(1+\alpha)}t^\alpha\;,
\end{equation}
for $0<\alpha<1$, for $\alpha=1$, one deals with normal diffusion. Subdiffusion is non-Markovian stochastic process, different from normal diffusion. However, as was shown in \cite{dybiec}, there is a non-Markovian process which provides the relation (\ref{eq1}) in which $\alpha=1$. Thus, it seems to be a good idea to include a stochastic interpretation in the definition of anomalous diffusion together with the relation (\ref{eq1}). Such a simple interpretation has a random walk, which is used in our considerations. We mention here that the random walk model is universal, for example, it has been used to derive normal diffusion--reaction equations \cite{rodriguez,abramson,helman} or subdiffusion--reaction equations \cite{zoia,henry,yadav,eule,hw}. 

The most commonly used differential equation described anomalous diffusion is the following equation with the Riemann--Liouville fractional derivative 
\begin{equation}\label{eq2}
\frac{\partial}{\partial t}P(x,t)=D_\alpha\frac{\partial^{1-\alpha}_{RL}}{\partial t^{1-\alpha}}\frac{\partial^2}{\partial x^2}P(x,t)\;,
\end{equation}
the derivative is defined for $\alpha>0$ as follows \cite{oldham,podlubny}
\begin{equation}\label{eq3}
\frac{d^\alpha_{RL}}{dt^\alpha}f(t)=\frac{1}{\Gamma(n-\alpha)}\frac{d^n}{dt^n}\int_0^t{(t-t')^{n-\alpha-1}f(t')dt'}\;,
\end{equation}
$n$ is a natural number fulfilled $\alpha\leq n<\alpha+1$. Equation (\ref{eq2}) was derived within the continuous time random walk formalism \cite{compte1996,mk}. Equation (\ref{eq2}) can be transformed to the following equation with the Caputo fractional derivative (see Eqs. (\ref{b10}) and (\ref{b11}), Appendix A)
\begin{equation}\label{eq4}
\frac{\partial^\alpha_C}{\partial t^\alpha}P(x,t)=D_\alpha\frac{\partial^2}{\partial x^2}P(x,t)\;,
\end{equation}
where \cite{podlubny}
\begin{equation}\label{eq5}
\frac{d^\alpha_C}{dt^\alpha}f(t)=\frac{1}{\Gamma(n-\alpha)}\int_0^t{(t-t')^{n-\alpha-1}\frac{d^n}{dt'^n}f(t')dt'}\;,
\end{equation}
$n-1<\alpha\leq n$.
The fundamental solution to Eqs. (\ref{eq2}) and (\ref{eq4}) (the Green function), which is defined by its initial condition $P(x,t;x_0)=\delta_{x,x_0}$ (in the following $\delta_{x,x_0}$ denotes both the Dirac--delta function for continuous variables and the Kronecker symbol for discrete ones), is interpreted as a probability density to find a random walker at point $x$ after time $t$ under the condition that its initial position is $x_0$. However, it is well known that the Green function of Eq. (\ref{eq2}) has a non--physical property. Namely, it has non--zero values for any $x$ at $t>0$. This means that some of the particles moves with an arbitrarily chosen large velocity. To avoid this absurdity the persistent random walk model was proposed \cite{cattaneo,haus,weiss}. Under the assumption that the actual random walker's step is correlated with the previous one, which means that the direction of successive steps is kept with some probability, for the normal diffusion process, one obtains the following differential hyperbolic Cattaneo equation 
\begin{equation}\label{eq6}
\tau \frac{\partial^2}{\partial t^2}P(x,t)+\frac{\partial}{\partial t}P(x,t)=D\frac{\partial^2}{\partial x^2}P(x,t)\;,
\end{equation}
where $D$ is the normal diffusion coefficient. A solution to this equation is above zero in a finite domain only, which ensures that a random walker's velocity is limited. We mention here that one of the simplest interpretations of this process is that the probability flux is delayed over time by parameter $\tau$ with respect to the probability gradient,
$J(x,t+\tau)=-D\frac{\partial}{\partial x}P(x,t)$.
Assuming $\tau\ll t$ there is 
$J(x,t)+\tau\frac{\partial}{\partial t}J(x,t)=-D\frac{\partial}{\partial x}P(x,t)$.
Combining the above equation with the continuity equation
$\frac{\partial}{\partial t}P(x,t)=-\frac{\partial}{\partial x}J(x,t)$,
one obtains Eq. (\ref{eq6}). For $\tau=0$ we have the normal diffusion equation. The generalization of Eq. (\ref{eq6}) to the subdiffusion system is not obvious. As was discussed in \cite{compte}, there are various forms of a such generalization, which are not equivalent to each other. 

The situation is more complicated when diffusing particles of species $A$ and $B$ can chemically react with each other according to the formula $n_A A+n_B B\rightarrow \emptyset(inert)$. Phenomenologically, the diffusion--reaction equations are derived on the basic of a normal diffusion equation (without the persistent effect, $\tau=0$) by subtracting a reaction term $\Pi(C_A,C_B)$ from the right--hand side of Eq. (\ref{eq6}). Within the mean--field approximation the reaction term reads \cite{bah}
\begin{equation}\label{eq7}
\Pi(C_A,C_B)=kC_A^{n_A}C_B^{n_B}\;,
\end{equation}
where $k$ is a reaction rate, $C_{A,B}$ denotes substance concentrations. Such a procedure provides the standard normal diffusion--reaction equation
\begin{equation}\label{eq8}
\frac{\partial}{\partial t}C_{i}(x,t)=D_i\frac{\partial^2}{\partial x^2}C_{i}(x,t)-n_i kC_A^{n_A}C_B^{n_B}\;,
\end{equation}
where $i=A,B$. A similar procedure was applied to obtain a subdiffusion--reaction equation. However, there arose a problem concerning which of the subdiffusion equations (\ref{eq2}) or (\ref{eq4}) should be taken into account. In \cite{sung} the reaction term was subtracted from the right--hand side of Eq. (\ref{eq2}), whereas in \cite{seki} it was shown that this term should be subtracted from the right--hand side of Eq. (\ref{eq4}). The above mentioned versions of subdiffusion--reaction equations describe the processes which dynamics differ from each other (see Appendix B). The latter version of the equation has been considered in many papers, for example in \cite{yuste,kosztlew}. We mention here that another version of the subdiffusion--reaction equation was derived in \cite{sokolov}.

The character of transport processes (normal diffusion or subdiffusion) strongly influences the dynamics of chemical reactions \cite{seki,s}. There arises a question concerning the influence of the persistent random walk effect on the subdiffusion--reaction process. In some physical systems this effect plays an important role. For example, as we showed in \cite{kosztlew1}, in electrochemical systems the Nyquist plots of subdiffusive impedance strongly depend on parameter $\alpha$ as well as on the parameter described by persistent random walk effect. A similar effect can occur in a system in which subdiffusive particles of species $A$ can chemically react with particles $B$. The reason is that the reaction efficiency depends on the particle's concentration. The probability that the reaction between particles $A$ and $B$, which are located close to each other, occurs in some time interval strongly depends on the character of particles' transport mechanism \cite{seki}. Moreover, as we will discuss later, the reaction rate for the persistent random walk is changed compared to the non--persistent one. 

In our paper we derive a Cattaneo--type subdiffusion--reaction equation which describes the persistent subdiffusive random walk with a chemical reaction of type $A+B\rightarrow B$. We assume that the three--dimensional system is homogeneous in the plane perpendicular to the $x$ axis, so it can be treated as a one--dimensional system. The particles $B$ are assumed to be immobile and all of them are located at the position $x_r$. In practice, this problem can be treated as a particle's random walk on a lattice with a single immobile trap. This system was chosen for theoretical study for the following reasons. Firstly, the concentration of particles $B$ does not change over time, thus the analytical treatment of the problem is relatively simpler than for other systems. The results obtained can be treated as the background to finding a more general equation for the case of chemical reactions $n_A A+n_B B\rightarrow\emptyset$ in a system with both mobile $A$ and $B$ species. Secondly, the model can be used for a theoretical description of the process in a system in which the reaction is ruled according to $A+B\rightarrow \emptyset$ if the concentration of static particles $B$ (located at the permeable membrane) remains `almost constant', which is achieved if the concentration is very large compared to the concentration of particles $A$ \cite{kosztlew2}. Such a model can be useful to describe  transport in a porous medium with a chemical reaction occurring at the medium surface \cite{parada}. Moreover, the time evolution of the concentration of $A$ particles can be measured experimentally, for example by means of the laser interferometric method \cite{kdm}, which gives the possibility of the experimental verification of the theoretical subdiffusion--reaction model. We add that the experimental method of concentration measurement mentioned above is effective for the (sub)diffusion--reaction systems with only one mobile substance.

The paper is organized as follows. In Sec.II we consider a non-persistent random walk in a system in which a particle $A$ can be absorbed with some probability into an arbitrary chosen site (this situation corresponds to the reaction $A+B\rightarrow B$ occurring in this site). Starting from difference equations with discrete time and space variables, we will derive the fractional subdiffusion--reaction equation for continuous variables. Next, we will generalize the obtained equation to the case of mobile $A$ and $B$ particles, which can chemically react according to a more complicated rule. The main aim of this section is to check if the method used in this paper provides the subdiffusion--reaction equation which was derived in \cite {seki} for the non--persistent random walk model. In Sec.III we will use a procedure to find the subdiffusion--reaction equation within the persistent random walk model. We will also find the solution to the equation over a long time limit for the case of the reaction $A+B(static)\rightarrow B(static)$. The discussion of various aspects of the model will be presented in Sec.IV. The details of the calculations and some useful formulae will be presented in three Appendixes.

\section{Subdiffusion--reaction equation}

We consider the non--persistent random walk in a discrete system in which a random walker $A$ can react with a static particle $B$ located at $m_r$, according to the formula $A+B\rightarrow B$. The subdiffusion--reaction equation was derived already using the random walk model with a continuous time, and here we will show that the lattice random walk model with discrete time (which is represented by the number of steps) provides the subdiffusion--reaction equation equivalent to the one derived in \cite{seki}. 

\subsection{General equation}

Let $P_n(m)$ denote a probability of finding a particle $A$ which arrives at site $m$ at the $n$--th step. If a particle arrives at the site $m_r$ then it can react during its stay at site $m_r$ with a particle $B$ with the probability $R$. This process is described by the following difference equation 
\begin{eqnarray}\label{eq9}
P_{n+1}(m;m_0)&=&\frac{1}{2}P_n(m+1;m_0)+\frac{1}{2}P_n(m-1;m_0)\nonumber\\
&-&\delta_{m,m_r}R P_n(m;m_0)\;,
\end{eqnarray}
$m_0$ is the initial position of the particle, $P_0(m)=\delta_{m,m_0}$. This equation is usually solved by means of the generating function method \cite{montroll64}. The generating function is defined as 
\begin{equation}\label{eq10}
S(m,z;m_0)=\sum_{n=0}^\infty{z^n P_n(m;m_0)}\;.
\end{equation}
From Eqs. (\ref{eq9}) and (\ref{eq10}) we obtain
\begin{eqnarray}\label{eq11}
S(m;m_0)-P_0(m;m_0)=\frac{z}{2}S(m+1;m_0)\\
+\frac{z}{2}S(m-1;m_0)-z\delta_{m,m_r}RS(m;m_0)\;,\nonumber
\end{eqnarray}

The probability of finding the particle at site $m$ for a continuous time equals $P(m,t;m_0)=\sum_{n=0}^\infty P_n(m;m_0)\Phi_n(t)$, where $\Phi_n(t)$ is the probability that a particle, starting from $m_0$, reaches site $m$ in $n$ steps. The function $\Phi$ depends on the waiting time probability density $\omega(t)$ needed to take the particle's next step. In terms of the Laplace transform, $L\{f(t)\}\equiv \hat{f}(s)=\int_0^\infty{{\rm exp}(-st)f(t)dt}$, one obtains \cite{mk} 
$\hat{\Phi}(s)=[1-\hat{\omega}(s)]\hat{\omega}^n(s)/s$,
which, together with Eq. (\ref{eq10}) provides
\begin{equation}\label{eq12}
\hat{P}(m,s;m_0)=\frac{1-\hat{\omega}(s)}{s}S(m,\hat{\omega}(s);m_0)\;.
\end{equation}

When a particle $A$ reaches the site $m_r$, it can react with a particle $B$. Let us assume that the distribution function of the reaction is $\psi(t)=\gamma{\rm exp}(-\gamma t)$, where $\gamma$ is the reaction rate. The waiting time distribution that the reaction will produce is as follows
\begin{equation}\label{eq13}
\psi_r(t)=\gamma {\rm exp}(-\gamma t)\left[1-\int_0^t{\omega(t')dt'}\right]\;,
\end{equation}
the last term of the right--hand side of the above equation (in the square bracket) represents the probability that the particle does not change its position in the time interval $(0,t)$. The probability that the particle reacts with single particle $B$ equals
\begin{equation}\label{eq14}
R=\int_0^\infty{\psi_r(t')dt'}=1-\hat{\omega(\gamma)}\;.
\end{equation}
Let us assume that the distance between discrete sites equals $\Delta x$. To pass from a discrete to a continuous space variable one puts $x=m\Delta x$, $P(m,t;m_0)=(\Delta x) P(x,t;x_0)$ and assumes that $\Delta x$ goes to zero. Taking into account the following approximation 
\begin{eqnarray}\label{eq15}
P(x\pm\Delta x,t;x_0)&=&P(x,t;x_0) \pm (\Delta x)\frac{\partial}{\partial x}P(x,t;x_0)\nonumber\\&+&\frac{(\Delta x)^2}{2}\frac{\partial^2}{\partial x^2}P(x,t;x_0)\;,
\end{eqnarray}
and Eqs. (\ref{eq11})-(\ref{eq15}), putting $z=\hat{\omega}(s)$, we obtain the following equation
\begin{eqnarray}\label{eq16}
& &\left[1-\hat{\omega}(s)\right]\hat{P}(x,s;x_0)-\frac{1-\hat{\omega}(s)}{s}P(x,0;x_0)\nonumber\\
&=&\hat{\omega}(s)\frac{(\Delta x)^2}{2}\frac{\partial^2}{\partial x^2}P(x,t;x_0)\\&-&\delta_{x,x_r}[1-\hat{\omega}(\gamma)]\hat{\omega}(s)\hat{P}(x,s;x_0)\nonumber\;.
\end{eqnarray}
The above equation, written in terms of Laplace transform, is the basis for further considerations. In the next subsection we will find the equation for a continuous time variable.

\subsection{Continuous time random walk approach}

Subdiffusion can be interpreted as a random walk in which the mean waiting time between a particle's successive steps is infinite whereas all the moments of the step length distribution are finite. We choose $\omega$ as the one--sided $\alpha$--stable distribution, which Laplace transform reads (here $0<\alpha<1$) \cite{hughes}
\begin{equation}\label{eq17}
\hat{\omega}(s)={\rm exp}(-\tau_\alpha s^\alpha)\;.
\end{equation}
Within the continuous time random walk formalism, function (\ref{eq17}) is considered in a limit of small $s$, which, according to the Tauberian theorems, corresponds to the limit of a large time $t$ \cite{hughes}, is
\begin{equation}\label{eq18}
\hat{\omega}(s)\approx 1-\tau_\alpha s^\alpha\;.
\end{equation}
In our approach, the length of the particle's step is not a random variable, but we can choose the parameter $\Delta x$ in such a way that the coefficient $(\Delta x)^2/2$ equals the variation of step length distribution which can be involved into a stochastic model. The definition of the subdiffusion coefficient then reads
\begin{equation}\label{eq19}
D_\alpha=\frac{(\Delta x)^2}{2\tau_\alpha}\;.
\end{equation}
The parameters $\alpha$ and $D_\alpha$ control subdiffusion and are measured experimentally \cite{kdm}. The parameter $\Delta x$ is related to $\tau_\alpha$ by Eq. (\ref{eq19}). Thus, we consider both $\Delta x$ and $\tau_\alpha$ as `small parameters'. From Eqs. (\ref{eq14}) and (\ref{eq18}), in the limit of small $\tau_\alpha$ we obtain $\tilde{R}\equiv R/\tau_\alpha=\gamma^\alpha$. From Eqs. (\ref{eq16}), (\ref{eq18}) and (\ref{eq19}), keeping the terms of the first order with respect to $\tau_\alpha$, after simple calculations we obtain
\begin{equation}\label{eq20}
\frac{\partial^\alpha_C}{\partial t^\alpha}P(x,t;x_0)=D_\alpha\frac{\partial^2}{\partial x^2}P(x,t;x_0)-\tilde{R} P(x,t;x_0)\;.
\end{equation}

Let us generalize Eq. (\ref{eq20}) to a system containing a large number of particles $A$ and $B$. Using the relation $C_A(x,t)=N_A P(x,t)$, where $C_A$ is the concentration of the particles $A$, $N_A$ denotes the initial number of particles $A$ in the system ($N_A\gg 1$), we find that Eq. (\ref{eq20}) is fulfilled by concentration $C_A$ and the reaction term reads $\Pi(C_A)=\tilde{R} C_A(x,t)$. The function $\tilde{R}$ is now proportional to the concentration of $B$ particles, thus we obtain $\Pi(C_A,C_B)=\gamma^\alpha C_A(x,t)C_B(x_r)$. The essential assumption is that the above reaction term is correct in the case of mobile particles $B$, which react with particle $A$ according to the more general formula $n_A A+n_B B\rightarrow \emptyset$. In general, within the mean field approach the reaction term is proportional to the probability that $n_A$ particles of species $A$ and $n_B$ particles of species $B$ meet in such a small volume that chemical reaction is possible with a probability controlled by the reaction rate $k$ (in the considerations presented above we have $k=\gamma^\alpha$). Thus, the arguments presented above suggest that the following equation  
\begin{equation}\label{eq21}
\frac{\partial^\alpha_C}{\partial t^\alpha}C_i(x,t)=D_\alpha\frac{\partial^2}{\partial x^2}C_i(x,t)-n_i\Pi(C_A,C_B)\;,
\end{equation}
$i=A,B$, where the reaction term is given by Eq. (\ref{eq7}), is the generalization of Eq. (\ref{eq20})

\section{Cattaneo--type subdiffusion--reaction equation}

\subsection{General equation}

Let $P^+_n(m)$, $P^-_n(m)$ denote probabilities that the particles arrive at site $m$ at step $n$ with a positive or negative velocity, respectively (in the following we will skip the symbol $m_0$ for shortening the notation), $\beta$ is a probability that a particle changes its velocity sense after arriving at site $m$. The persistent random walk with a reaction with a particle $B$ located at $m_r$ is described by the following equations \cite{weiss} 
\begin{eqnarray}
P_{n+1}^+(m)&=&(1-\beta)P_n^+(m+1)+\beta P_n^-(m+1)\nonumber\\&-&R_\beta P^+_n(m)\delta_{m,m_r}\;,\label{eq22}\\
P_{n+1}^-(m)&=&(1-\beta)P_n^-(m-1)+\beta P_n^+(m-1)\nonumber\\&-&R_\beta P^-_n(m)\delta_{m,m_r}\;.\label{eq23}
\end{eqnarray}
As in the previous section, we solve the equations by means of the generating function method. Generally, the reaction coefficient $R$ depends on parameter $\beta$. We can motivate this fact as follows. Various theoretical reaction models, applied to the reaction $A+B(static)\rightarrow B(static)$, assume that the particle $A$ overcome the potential barrier during its movement along the reaction coordinate axis \cite{kramers,hanggi}. However, the probability of passing the barrier depends on a particle's energy. If the particle comes to the site $m_r$ and its velocity sense is not changed after arriving at site, we assume that the reaction occurs with reaction rate $\gamma_1$, but if the particle's velocity sense is changed, the energy of the particle will be lower than in the previous case, thus the reaction occurs with the reaction rate $\gamma_2$, $\gamma_2<\gamma_1$. The probability of the `choice' of the reaction rate depends on $\beta$. Thus, we postulate that
\begin{equation}\label{eq24}
R_\beta=(1-\beta)[1-\hat{\omega}(\gamma_1)]+\beta[1-\hat{\omega}(\gamma_2)]\;.
\end{equation}

\begin{figure}[!ht]
\centering
\includegraphics[height=5.7cm]{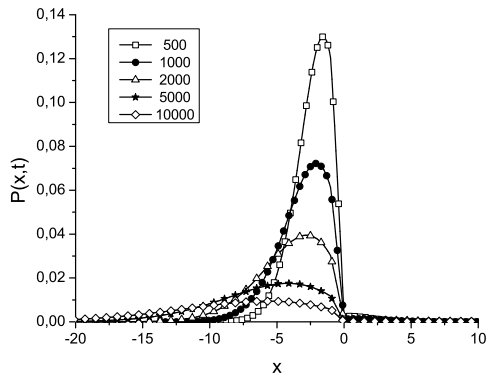}
\caption{Functions (\ref{eq38}) for various values of time $t$ given in the legend, $D_\alpha=0.01$, $\alpha=0.9$, $\beta=0.6$, $\gamma_1=0.5$, $\gamma_2=0.2$, $x_0=-1$, $x_r=0$ (all quantities are given in arbitrary chosen units).}\label{Fig1}
\end{figure}

\begin{figure}[!ht]
\centering
\includegraphics[height=5.7cm]{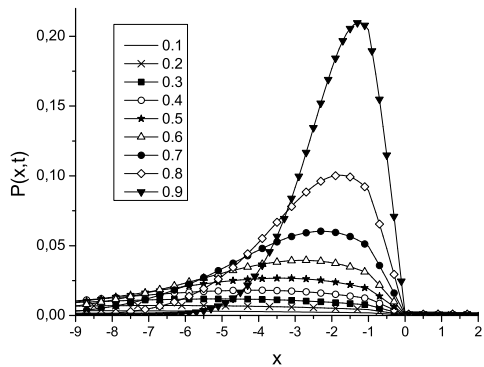}
\caption{Functions (\ref{eq38}) for various values of probability $\beta$ given in the legend, $t=2000$, the other parameters are the same as in Fig.1.}\label{Fig2}
\end{figure}

Proceeding similarly as in the previous case, Eqs. (\ref{eq22}) and (\ref{eq23}), transformed to the continuous variables $(x,t)$, in terms of Laplace transform read
\begin{widetext}
\begin{eqnarray}
\hat{P}^+(x,s)-\frac{1-\hat{\omega}(s)}{s}P^+(x,0)&=&\hat{\omega}(s)\left\{(1-\beta)\left[\hat{P}^+(x,t)-(\Delta x)\frac{\partial}{\partial x}\hat{P}^+(x,s)+\frac{(\Delta x)^2}{2}\frac{\partial^2}{\partial x^2}\hat{P}^+(x,s)\right]\right.\label{eq25}\\
&+&\left.\beta\left[\hat{P}^-(x,t)-(\Delta x)\frac{\partial}{\partial x}\hat{P}^-(x,s)+\frac{(\Delta x)^2}{2}\frac{\partial^2}{\partial x^2}\hat{P}^-(x,s)\right] \right\}-R_\beta\hat{P}^+(x,s)\delta_{x,x_r}\;,\nonumber\\
\nonumber\\
\hat{P}^-(x,s)-\frac{1-\hat{\omega}(s)}{s}P^-(x,0)&=&\hat{\omega}(s)\left\{(1-\beta)\left[\hat{P}^-(x,t)+(\Delta x)\frac{\partial}{\partial x}\hat{P}^-(x,s)+\frac{(\Delta x)^2}{2}\frac{\partial^2}{\partial x^2}\hat{P}^-(x,s)\right]\right.\label{eq26}\\
&+&\left.\beta\left[\hat{P}^+(x,t)+(\Delta x)\frac{\partial}{\partial x}\hat{P}^+(x,s)+\frac{(\Delta x)^2}{2}\frac{\partial^2}{\partial x^2}\hat{P}^+(x,s)\right] \right\}-R_\beta\hat{P}^-(x,s)\delta_{x,x_r}\;.\nonumber
\end{eqnarray}
\end{widetext}
The probability density of finding a particle at site $x$ is
\begin{equation}\label{eq27}
\hat{P}(x,s)=\hat{P}^+(x,s)+\hat{P}^-(x,s)\;.
\end{equation}
Let us define the flux as follows
\begin{equation}\label{eq28}
\hat{J}(x,s)=\hat{J}^+(x,s)-\hat{J}^-(x,s)\;.
\end{equation}
Adding and next subtracting Eqs. (\ref{eq25}) and (\ref{eq26}), taking into account (\ref{eq27}) and (\ref{eq28}), we obtain
\begin{eqnarray}\label{eq29}
& &[1-\hat{\omega}(s)]\hat{P}(x,s)-\frac{1-\hat{\omega}(s)}{s}P(x,0)\\&=& \hat{\omega}(s)\left[\frac{(\Delta x)^2}{2}\frac{\partial^2}{\partial x^2}\hat{P}(x,s)-(1-2\beta)(\Delta x)\frac{\partial}{\partial x}\hat{J}(x,s)\right]\nonumber\\&-&\hat{\omega}(s)\delta_{x,x_r}R_\beta\hat{P}(x,s)\nonumber
\nonumber\;,
\end{eqnarray}
and
\begin{eqnarray}\label{eq30}
& &[1-(1-2\beta)\hat{\omega}(s)]\hat{J}(x,s)-\frac{1-\hat{\omega}(s)}{s}J(x,0)\nonumber\\&=& -\hat{\omega}(s)(\Delta x)\frac{\partial}{\partial x}\hat{P}(x,s)+(1-2\beta)\hat{\omega}(s)\\
&\times& \frac{(\Delta x)^2}{2}\frac{\partial^2}{\partial x^2}\hat{J}(x,s)-\hat{\omega}(s)\delta_{x,x_r}R_\beta\hat{J}(x,s)\nonumber\;.
\end{eqnarray}
The assumption that the probability of the taking of a particle's first step with negative velocity is equal to the probability of the taking of a particle's first step with positive velocity gives $J(x,0)=0$. Combining equations (\ref{eq29}) and (\ref{eq30}) we obtain the following equation, which is the base for the derivation of differential subdiffusion--reaction equations for various probability densities $\omega$ 
\begin{widetext}
\begin{eqnarray}\label{eq31}
& &[1-\hat{\omega}(s)][1-(1-2\beta)\hat{\omega}(s)]\left[\hat{P}(x,s)-\frac{P_0(x)}{s}\right]
=\hat{\omega}(s)[1+(1-2\beta)\hat{\omega}(s)]\frac{(\Delta x)^2}{2}\frac{\partial^2}{\partial x^2}\hat{P}(x,s)\nonumber\\
&+&(1-2\beta)\hat{\omega}(s)[1-\hat{\omega}(s)]\frac{(\Delta x)^2}{2}\frac{\partial^2}{\partial x^2}\left[\hat{P}(x,s)-\frac{P_0(x)}{s}\right]
-R_\beta\delta_{x,x_r}\left\{[1-\hat{\omega}(s)]\hat{\omega}(s)\left[\hat{P}(x,s)-\frac{P_0(x)}{s}\right]\right.\\
&-&\left.[\hat{\omega}(s)(1-\hat{\omega}(s))+(1-2\beta)\hat{\omega}(s)]\hat{P}(x,s)
- [1+(1-2\beta)]\hat{\omega}^2(s)\frac{(\Delta x)^2}{2}\frac{\partial^2}{\partial x^2}\hat{P}(x,s)\right\}\;.\nonumber
\end{eqnarray}
\end{widetext}

\subsection{Continuous time random walk approach}

Parameter $\beta$ controls the correlation of jumps, namely the correlation coefficient is $cor=\left\langle (\Delta x)_n(\Delta x)_{n+1}\right\rangle=(1-2\beta)(\Delta x)^2$, where $(\Delta x)_n$ is the particle's displacement during its $n$--th step \cite{haus,pottier}. Thus, the case of $\beta=1/2$ corresponds to the `ordinary' non--persistent random walk, described by the (sub)diffusion equation. In the various forms of the Cattaneo subdiffusion equation a parameter analogous to $\tau$ occurring in (\ref{eq6}) is present (in the following we denote this parameter by $\tilde{\tau}_\alpha$). Motivated by the above mentioned facts, we assume that all terms containing $\tilde{\tau}_\alpha$ in the Cattaneo subdiffusion--reaction equation should vanish if $\beta=1/2$ and the equation obtained take the form of a `standard' subdiffusion--reaction equation (\ref{eq21}) with the reaction term (\ref{eq7}). Moreover, in different versions of the Cattaneo subdiffusion equation which have been considered until now the terms of the order $\Theta(\tilde{\tau}_\alpha^2)$ do not occur \cite{compte}. In order to derive a new equation from (\ref{eq31}), which fulfils the above conditions, we set the following rules:
\begin{enumerate}
	\item the approximation of the function $\hat{\omega}(s)$ is given by Eq. (\ref{eq18}),
	\item the parameters $\alpha$ and $D_\alpha$ (the last one is defined by Eq. (\ref{eq19})) are the same for both persistent and non--persistent models, moreover
	\begin{equation}\label{eq32}
	(\Delta x)^2=2D_\alpha \tau_\alpha\;,
	\end{equation}
	\item according to Eq. (\ref{eq18}) and (\ref{eq24}), for small $\tau_\alpha$ 
	\begin{equation}\label{eq33}
	R_\beta =\tau_\alpha\tilde{R}_\beta\;,
	\end{equation}
	where $\tilde{R}_\beta=(1-\beta)\gamma_1^\alpha+\beta\gamma_2^\alpha$,
	\item in the obtained equation, we keep all terms up to the first order with respect to $\tau_\alpha$, the terms of the second order with respect to $\tau_\alpha$ are kept only in terms which vanish at $\beta=1/2$. The other terms are neglected.	
\end{enumerate}
Taking into account the above points, using inverse Laplace transforms (\ref{b4}), (\ref{b5}) (Appendix A) and  the following formula (here $0<\alpha<1$) \cite{podlubny}
\begin{equation}\label{eq34}
\frac{\partial^\alpha_{RL}}{\partial t^\alpha}P(x,t)=\frac{\partial^\alpha_C}{\partial t^\alpha}P(x,t)+\frac{t^{-\alpha}}{\Gamma(1-\alpha)}P(x,0)\;,
\end{equation}
we obtain from Eq. (\ref{eq31}) the following subdiffusion--reaction equation
\begin{widetext}
\begin{eqnarray}\label{eq35}
& &(1-2\beta)\tau_\alpha \frac{\partial^{2\alpha}_C}{\partial t^{2\alpha}}P(x,t)+2\beta \frac{\partial^\alpha_C}{\partial t^\alpha}P(x,t)
=2(1-\beta)D_\alpha\frac{\partial^2}{\partial x^2}P(x,t)
-(1-2\beta)\tau_\alpha D_\alpha\frac{\partial^2}{\partial x^2}\frac{\partial^\alpha_C}{\partial t^\alpha}P(x,t)\\
&-&\tilde{R}_\beta\delta_{x,x_r}\left[2\beta P(x,t)+2(1-2\beta)\tau_\alpha \frac{\partial^\alpha_C}{\partial t^\alpha}P(x,t)+2\tau_\alpha\frac{t^{-\alpha}}{\Gamma(1-\alpha)}P(x,0)
-(1-2\beta)\tau_\alpha D_\alpha \frac{\partial^2}{\partial x^2}P(x,t)\right]\;.\nonumber
\end{eqnarray}
\end{widetext}

\subsection{The solution}

The general form of the solution to Eq. (\ref{eq35}) in terms of Laplace and Fourier transforms is given in the Appendix C, Eq. (\ref{a1}). Over a long time limit, the solution given in terms of Laplace transform reads (here $\tilde{R}\neq 0$)
\begin{eqnarray}\label{eq36}
\hat{P}(x,s)&=&\frac{s^{-1+\alpha/2}}{2\sqrt{\tilde{D}_\alpha}}\;{\rm exp}\left(-\frac{|x-x_0|s^{\alpha/2}}{\sqrt{\tilde{D}_\alpha}}\right)\\
&-&\frac{s^{-1+\alpha/2}}{2\sqrt{\tilde{D}_\alpha}}\;\frac{{\rm exp}\left(-\frac{(|x|+|x_0|)s^{\alpha/2}}{\sqrt{\tilde{D}_\alpha}}\right)}{1+(2\sqrt{\tilde{D}_\alpha}s^{\alpha/2}/\tilde{R})}\nonumber\;.
\end{eqnarray}
Using the following formula \cite{koszt}
\begin{eqnarray}\label{eq37}
L^{-1}\{s^\nu{\rm exp}(-as^\gamma)\}\equiv f_{\nu,\gamma}(t;a)\\=\frac{1}{t^{\nu+1}}\sum_{k=0}^\infty{\frac{1}{k!\Gamma(-k\gamma-\nu)}\left(-\frac{a}{t^\gamma}\right)}\nonumber\;,
\end{eqnarray}
$a,\gamma>0$, we obtain
\begin{eqnarray}\label{eq38}
P(x,t)&=&\frac{1}{2\sqrt{\tilde{D}_\alpha}}f_{-1+\alpha/2,\alpha/2}\left(t;\frac{|x-x_0|}{\sqrt{\tilde{D}_\alpha}}\right)\nonumber\\
&-&\frac{1}{2\sqrt{\tilde{D}_\alpha}}\sum_{k=0}^\infty\left(-\frac{2\sqrt{\tilde{D}_\alpha}}{\tilde{R}_\beta}\right)^k\\
&\times& f_{-1+(k+1)\alpha/2,\alpha/2}\left(t;\frac{|x|+|x_0|}{\sqrt{\tilde{D}_\alpha}}\right)\;,\nonumber
\end{eqnarray}
where 
\begin{equation}\label{eq39}
\tilde{D}_\alpha=\frac{1-\beta}{\beta}D_\alpha\;.
\end{equation}
We add that the mathematical condition of a long time limit is briefly described in Appendix A, in the comment just after Eq. (\ref{b6}).
In Fig.1 and Fig.2 there are presented example plots of function (\ref{eq38}). Fig.1 shows that the solutions to Eq. (\ref{eq35}) over the long time limit behave `almost' in the same way as for the system with an absorbing wall located at $x_r$. The plots presented in Fig.2 show that the solutions strongly depend on parameter $\beta$.

\subsection{More general form of subdiffusion--reaction equation}

Let us generalize Eq. (\ref{eq35}) to a many--particle system containing substances $A$ and $B$. The generalization is based on the interpretation of the subdiffusion--reaction process. If particles $A$ move independently of each other, the particle's concentration, defined as
$C_A(x,t)=N_A P(x,t)$,
also fulfils Eq. (\ref{eq35}), but now the reaction probability $\tilde{R}_\beta$ depends on the concentration of particles $B$. For the reaction $A+B(static)\rightarrow B(static)$, if all particles $B$ are located at $x_r$, then $\tilde{R}_\beta=kC_B(x_r)$.
To simplify the description let us introduce the function $\Psi$, whose value is proportional to the probability of meeting particles $A$ and $B$ in such a small volume that a chemical reaction is possible; the probability of the reaction is then controlled by the reaction rate $k$. For the considered reaction we have
\begin{equation}\label{eq40}
\Pi(C_A,C_B)=k\Psi(x,t)\;,
\end{equation}
with $\Psi(x,t)=C_A(x,t)C_B(x_r)$. 
When particles $B$ are mobile, the crucial assumption is that Eq. (\ref{eq40}) is still valid and 
\begin{equation}\label{eq41}
\Psi(x,t)=C_A(x,t)C_B(x,t)\;.
\end{equation}
Equation (\ref{eq41}) is also assumed to be valid for the reaction $A+B\rightarrow \emptyset$.

The parameters $D_\alpha$, $\beta$ and $\tau_\alpha$ are assumed to be defined separately for substances $A$ and $B$ whereas the parameter $\alpha$ is assumed to be the same for both substances. The last assumption is motivated by experiments \cite{kdm} which suggest that $\alpha$ is determined by the properties of a medium, whereas the others depend on properties of both the particles and medium. As suggested by Eq. (\ref{eq35}), for $\beta_{1,2}\neq 0$, the general form of the Cattaneo--type subdiffusion--reaction equation reads
\begin{eqnarray}\label{eq42}
\tilde{\tau}_{\alpha,i} \frac{\partial^{2\alpha}_C}{\partial t^{2\alpha}}C_i(x,t)+\frac{\partial^\alpha_C}{\partial t^\alpha}C_i(x,t)
=\tilde{D}_{\alpha,i}\frac{\partial^2}{\partial x^2}C_i(x,t)
\\-\tilde{\tau}_{\alpha,i} D_\alpha\frac{\partial^2}{\partial x^2}\frac{\partial^\alpha_C}{\partial t^\alpha}C_i(x,t)-\Pi_i[C_A,C_B]\;.\nonumber
\end{eqnarray}
where $i=A,B$, $\tilde{\tau}_{\alpha,i}=(1-2\beta_i)\tau_{\alpha,i}/(2\beta_i)$, $\tilde{D}_{\alpha,i}=(1-\beta_i)D_{\alpha,i}/\beta_i$, $C_i(x,t)=N_i P_i(x,t)$, $P_i(x,t)$ is the probability density of finding a particle of species $i$ at position $x$ and time $t$, $N_i$ denoting the initial number of particles $i$. 

There is a problem in finding a proper reaction term $\Pi$. It is not obvious which the position of $\tilde{R}_\beta\equiv\Psi/C_A$ (now depending on the variables $x$ and $t$) is within the reaction term occurring in (\ref{eq35}). More particularly, one should find out that the derivative operators act on the product $\tilde{R}_\beta C_A$ or on function $C_A$ alone. To solve this problem we recall the interpretation of the subdiffusion equation. 

Subdiffusion is a non--Markovian process generated by the anomalously long time of a particle's remaining in one position. On the other hand, it looks like some of the particles apparently temporarily `vanish', which means that they temporarily do not take part in the random walk process. This interpretation is supported by the phenomenological method of deriving the fractional Cattaneo equation. Namely, involving the fractional derivative into the flux equation 
$J(x,t)+\tau\frac{\partial_C^\alpha}{\partial t^\alpha}J(x,t)=-D\frac{\partial}{\partial x}P(x,t)$, and
combining the above equation with the fractional continuity equation
$\frac{\partial_C^\alpha}{\partial t^\alpha}P(x,t)=-\frac{\partial}{\partial x}J(x,t)$, we obtain a simplified form of the Cattaneo subdiffusion equation (\ref{eq35}) without chemical reactions. However, the fractional continuity equation does not accomplish the number of particles, which can be interpreted as follows.
The approximation of the fractional Caputo derivative (\ref{eq5}) reads \cite{kosztlew,podlubny,oldham}
\begin{eqnarray}\label{eq43}
\frac{\partial_C^\alpha}{\partial t^\alpha}C_A(x,t)=\frac{1}{(\Delta t)^\alpha}\Big[C_A(x,t)-C_A(x,t-\Delta t)\nonumber\\
-\sum_{k=1}^L{w_k C_A(x,t-k\Delta t)}-\frac{1}{t^\alpha\Gamma(1-\alpha)}C_A(x,0)\Big]\;,
\end{eqnarray}
where $w_1=\alpha-1$, $w_k=\alpha(1-\alpha)\ldots (k-1-\alpha)/k!$, $k\geq 2$, $L$ is the memory length.
The `apparently vanishing particles' effect is represented by a fractional derivative in the equation (\ref{eq35}) (more particularly, by the two last terms in square brackets in Eq. (\ref{eq43})) and provides a reduced effective concentration of particles which can be involved in chemical reactions. Thus, the subdiffusive effect regards the functions on which the fractional differential operator acts.

Let us return for a moment to Eq. (\ref{eq35}) putting $P(x,t)\rightarrow C_A(x,t)$, $\tilde{R}_\beta\rightarrow kC_B(x,t)$. If we assume that the second term in the square bracket on the right--hand side of this equation is in the form $kC_B(x,t)\partial_C^\alpha C_A(x,t)/\partial t^\alpha$, the subdiffusion effect does not concern the particles $B$ during the reaction process. The derivative of the second order with respect to $x$ can be approximated as $\partial^2 C_i(x,t)/\partial x^2\sim [C_i(x+\Delta x,t)+C_i(x-\Delta x,t)]/2-C_i(x,t)$. This term describes temporal changes in concentration generated by the concentration difference between the concentration measured in $x$ and the mean concentration measured in its vicinity, the velocity of this process is controlled by the subdiffusion coefficient. In this term, the memory effect is not present, but taking into account that a chemical reaction can be present at point $x$ as well as in its vicinity, we assume the following term occurring in Eq. (\ref{eq35}) $\partial^2 [C_A(x,t)C_B(x,t)]/\partial x^2$.
Summarizing the above considerations, in order to keep the subdiffusive effect in both substances we assume the following form of the reaction term
\begin{eqnarray}\label{eq44}
\Pi_i (C_A,C_B)=\Psi(x,t)+2\tilde{\tau}_{\alpha,i} \frac{\partial^\alpha_C}{\partial t^\alpha}\Psi(x,t)\\
-\tilde{\tau}_{\alpha,i} D_{\alpha,i} \frac{\partial^2}{\partial x^2}\Psi(x,t)+\frac{\tau_{\alpha,i}}{\beta_i}\frac{t^{-\alpha}}{\Gamma(1-\alpha)}\Psi(x,0)\;,\nonumber
\end{eqnarray}
where $\Psi(x,t)$ is given by Eq. (\ref{eq41}).

The generalization in the more general chemical reaction $n_A A+n_B B\rightarrow \emptyset(inert)$ seems to be natural within the mean field approximation and is given by the following equation
\begin{eqnarray}\label{eq45}
\tilde{\tau}_{\alpha,i} \frac{\partial^{2\alpha}_C}{\partial t^{2\alpha}}C_i(x,t)+\frac{\partial^\alpha_C}{\partial t^\alpha}C_i(x,t)
=\tilde{D}_{\alpha,i}\frac{\partial^2}{\partial x^2}C_i(x,t)
\\ -\tilde{\tau}_{\alpha,i} D_\alpha\frac{\partial^2}{\partial x^2}\frac{\partial^\alpha_C}{\partial t^\alpha}C_i(x,t)-n_i\Pi_i[C_A,C_B]\;,\nonumber
\end{eqnarray}
$i=A,B$, where the reaction term is given by Eq. (\ref{eq44}) with
\begin{equation}\label{eq46}
\Psi(x,t)=C^{n_A}_A(x,t)C^{n_B}_B(x,t)\;.
\end{equation}

\section{Final remarks}

The Cattaneo--type subdiffusion reaction equation (\ref{eq35}) for the reaction $A+B(static)\rightarrow B(static)$ was derived within the continuous time random walk formalism using the persistent random walk model, but its generalization in the cases of more complicated reactions was made using a `heuristic' method based on a stochastic interpretation of the subdiffusion--reaction process. Thus, Eqs. (\ref{eq42}) and (\ref{eq45}) should rather be treated as postulates. Nevertheless, we believe that this equation will be useful in modelling subdiffusion--reaction processes occurring in nature, since it has a relatively simple stochastic interpretation.

Let us note that parameter $\beta$ changes the effective subdiffusion coefficient. Namely, from Eq. (\ref{a2}) for the system without chemical reactions ($\tilde{R}_\beta=0$) we obtain 
\begin{equation}\label{eq47}
\left\langle (\Delta x)^2\right\rangle =\frac{2\tilde{D}_\alpha}{\Gamma(1+\alpha)}t^\alpha\;,
\end{equation}
where $\tilde{D}_\alpha$ is defined by (\ref{eq39}). We note that there are two definitions of subdiffusion coefficients. The first one, defined by Eq. (\ref{eq47}), shows how fast particles spread out over a long time limit (in this case every particle performs large number of steps); this coefficient we call the `effective subdiffusion coefficient'. It is obvious that it depends on parameter $\beta$. For example, if $\beta=1$, then a particle changes its velocity sense at every step with a probability of 1. In practice, the particle does not changes its position over time which provides $\tilde{D}_\alpha=0$. The second subdiffusion coefficient refers to a particle's single step and is defined by (\ref{eq19}) within the continuous time random walk formalism, and is independent of $\beta$. Both of them are equal to each other for $\beta=1/2$.

Persistent random walk is a process with memory, as well as subdiffusion. There arises a question: are these two effects simultaneously worth considering? The subdiffusive memory effect, controlled by the parameter $\alpha$, is long and vanishes in the case of normal diffusion. The persistent random walk memory effect, which arises from the correlation of the successive random walker's steps, is relatively short. This is controlled by parameter $\beta$, which is assumed to be independent of $\alpha$. As we can see in Fig. \ref{Fig2}, parameter $\beta$ significantly influences the solutions to Eq. (\ref{eq35}). The considerations presented in this paper show that the effect of step correlations changes the effective subdiffusion parameter $\tilde{D}_\alpha$ and provides new terms in the subdiffusion--reaction equation which can change the dynamic of the process (at least in some situations).

The dynamic of the process depends on parameter $\beta$.
For $\beta<1/2$ a particle prefers the direction of its previous step. This occurs when the particle inertia effect is taken into account. For $1/2<\beta<1$, one obtains the effect of the rapid changing of a particle's step direction which occurs more frequently than in the case of the uncorrelated random walk. This effect can be caused by the interaction of diffusing particles and it is expected to be in a system with large particles concentration. Diffusion or subdiffusion in dense systems, in which the effective diffusion coefficient depends on the concentration, is usually described by  non--linear equations, but we suppose that -- at least in some situations -- such a process can be described by the Cattaneo type subdiffusion equation with $\beta>1/2$.

The most simple approximation of the reaction term seems to be neglecting the terms occurring in (\ref{eq35}) which contain the small parameter $\tau_{\alpha,i}$. In this way, parameter $\beta$ will be involved in the reaction rate constant alone. However, by consequently neglecting similar terms in the remaining parts of the equation, we lose the steps' correlation effect. The reaction rate for the reaction $A+B(static)\rightarrow B(static)$ is given by Eq. (\ref{eq24}). However, this is the simplest situation in which the persistent random walk effect can be explicitly taken into account in derivation of the reaction rate coefficient. In the case of mobile $B$, and for more complicated reactions, the reaction rate cannot be defined in such a simple form. The generalization can be done using, for example, the diffusion model of chemical reactions described by difference--differential equations \cite{montroll67,mcquarrie}, in which rates depend on parameter $\beta$.

\begin{acknowledgments}
The author wants to thank Katarzyna D. Lewandowska for the revision of the manuscript.
This paper was partially supported by the Polish National Science Centre under grant No. 1956/B/H03/2011/40.
\end{acknowledgments}

\appendix

\section{Laplace transforms}

The Laplace transform of the Riemann--Liouville fractional derivative reads
\begin{equation}\label{b1}
L\left\{\frac{d_{RL}^\alpha f(t)}{dt^\alpha}\right\}=s^\alpha\hat{f}(s)-\sum_{k=0}^{n-1}s^k\left.\frac{d^{\alpha-k-1} f(t)}{dt^{\alpha-k-1}}\right|_{t=0}\;,
\end{equation}
$n-1\leq\alpha <n$,
where ($\gamma>0$)
\begin{equation}\label{b2}
\frac{d_{RL}^{-\gamma} f(t)}{dt^{-\gamma}}=\frac{1}{\Gamma(-\gamma)}\int_0^t (t-t')^{-\gamma-1}f(t')dt'\;.
\end{equation}
Let $0<\alpha<1$ and $f$ is bounded over the time interval $(0,t)$, $|f(t)|<A$, $t\in (0,t)$. Thus, 
\begin{eqnarray}\label{b3}
\left|\frac{d_{RL}^{\alpha-1} f(t)}{dt^{\alpha-1}}\right|<\frac{A}{\Gamma(\alpha-1)}\int_0^t (t-t')^{\alpha-1}dt'\\
=\frac{A t^\alpha}{\alpha\Gamma(\alpha-1)}\stackrel{t\rightarrow 0}{\rightarrow}0\nonumber\;.
\end{eqnarray}
Equations (\ref{b1})--(\ref{b3}) provide
\begin{equation}\label{b4}
L\left\{\frac{d_{RL}^\alpha f(t)}{dt^\alpha}\right\}=s^\alpha\hat{f}(s)\;.
\end{equation}
The above equation is also applied for the initial distribution function which is given formally by the delta--Dirac function, since this unbounded function is only an idealization of a realistic initial condition and can be approximated by a bounded one.

The Laplace transform of the Caputo fractional derivative reads
\begin{equation}\label{b5}
L\left\{\frac{d_C^\alpha f(t)}{dt^\alpha}\right\}=s^\alpha\hat{f}(s)-\sum_{k=0}^{n-1}s^{\alpha-k-1}\left.\frac{d^k f(t)}{dt^k}\right|_{t=0}\;,
\end{equation}
$n-1<\alpha\leq n$.

Using 
\begin{equation}\label{b9}
F\left\{\frac{\partial^2}{\partial x^2}P(x,t)\right\}=-k^2\hat{P}(k,t)\;,
\end{equation}
for $P(x,0)=\delta_{x,0}$ the Fourier and Laplace transforms of Eq. (\ref{eq2}) reads
\begin{equation}\label{b10}
s\hat{P}(k,s)-1=-s^{1-\alpha}k^2D_\alpha\hat{P}(k,s)\;.
\end{equation}
Transforming the above equation to the following form
\begin{equation}\label{b11}
s^\alpha\hat{P}(k,s)-s^{\alpha-1}=-k^2D_\alpha\hat{P}(k,s)\;,
\end{equation}
and using Eqs. (\ref{b5}) and (\ref{b9}) one obtains Eq. (\ref{eq4}).

Using Eq. (\ref{eq37}) and the exponent function ${\rm exp(u)}=\sum_{k=0}^\infty u^n/n!$, we obtain 
\begin{eqnarray}\label{b6}
& &L^{-1}\left\{s^\nu\sum_{k=0}^\infty\left(\frac{(-as^\gamma)^k}{k!}\right)\right\}\\
&=&\frac{1}{t^{\nu+1}}\sum_{k=0}^\infty \frac{1}{k!\Gamma(-\gamma k-\nu)}\left(-\frac{a}{t^\gamma}\right)^k\nonumber\;.
\end{eqnarray}
From (\ref{b6}) we obtain the condition $s\ll 1/a^{1/\gamma}$ corresponding to $t\gg a^{1/\gamma}$.

\section{Subdiffusion--reaction equations}

The subdiffusion--reaction equation can be obtained heuristically by subtracting the reaction term from the right--hand side of the subdiffusion equation. Putting $C_A(x,t)=N_A P_A(x,t)$ and $C_B(x,t)=N_B P_B(x,t)$, where $N_i$ denotes the initial number of particles of species $i$. Therefore, Eqs. (\ref{eq2}) and (\ref{eq4}) are fulfilled also by concentrations $C_A$ and $C_B$. In \cite{sung} the equation of the form was postulated
\begin{equation}\label{c1}
\frac{\partial}{\partial t}C_i(x,t)=D_\alpha\frac{\partial^{1-\alpha}_{RL}}{\partial t^{1-\alpha}}\frac{\partial^2}{\partial x^2}C_i(x,t)-\Pi(C_A,C_B)\;,
\end{equation}
$i=A,B$. Thus, subdiffusion does not influence the reaction process directly, since the reaction term is located outside the fractional Riemann--Liouville derivative.
 
In \cite{seki} the derivation of the subdiffusion--reaction equation provides following equation
\begin{equation}\label{c2}
\frac{\partial}{\partial t}C_i(x,t)=\frac{\partial^{1-\alpha}_{RL}}{\partial t^{1-\alpha}}\left[D_\alpha\frac{\partial^2}{\partial x^2}C_i(x,t)-\Pi(C_A,C_B)\right]\;.
\end{equation}
In this case the kinetic of reactions is controlled by parameter $\alpha$.
Eq. (\ref{c2}) can be transformed to 
\begin{equation}\label{c3}
\frac{\partial^\alpha_C}{\partial t^\alpha}C_i(x,t)=D_\alpha\frac{\partial^2}{\partial x^2}C_i(x,t)-\Pi(C_A,C_B)\;.
\end{equation}

\section{General solution of Eq. (\ref{eq35})}

The general solution of (\ref{eq35}) in terms of Laplace and Fourier transforms, $F\left\{f(x)\right\}\equiv\hat{f}(k)=\int_{-\infty}^\infty{\rm exp}(ikx)f(x)dx$, is
\begin{widetext}
\begin{equation}\label{a1}
\hat{P}(k,s)=\frac{(1-2\beta)\tau_\alpha s^{2\alpha-1}{\rm e}^{ikx_0}+2\beta s^{\alpha-1}{\rm e}^{ikx_0}-k^2(1-2\beta)\tau_\alpha s^{\alpha-1}-\tilde{R}_\beta[2\beta+2(1-2\beta)\tau_\alpha s^\alpha]\hat{P}(0,s) }{(1-2\beta)\tau_\alpha s^{2\alpha}+2\beta s^\alpha+k^2[2(1-\beta)D_\alpha-(1-2\beta)\tau_\alpha s^\alpha]}
\end{equation}
\end{widetext}
In the limit of small $s$ we obtain
\begin{equation}\label{a2}
\hat{P}(k,s)=\frac{s^{\alpha-1}{\rm e}^{ikx_0}-\tilde{R}_\beta\hat{P}(0,s) }{s^\alpha+k^2\tilde{D}_\alpha}
\end{equation}
where $\tilde{D}_\alpha=(1-\beta)D_\alpha/\beta$.
Using the following inverse Fourier transform ($a>0$) to Eq. (\ref{a2})
\begin{equation}\label{a3}
F^{-1}\left\{\frac{1}{a^2+k^2}\right\}=\frac{1}{2a}{\rm e}^{-a|x|}\;,
\end{equation}
we get
\begin{eqnarray}\label{a4}
\hat{P}(x,s)&=&\frac{1}{2\sqrt{\tilde{D}_\alpha s^\alpha}}\left[s^{\alpha-1}{\rm e}^{-\frac{|x-x_0|}{\sqrt{\tilde{D}_\alpha}}s^{\alpha/2}}\right.\\&-&\left.\tilde{R}\hat{P}(0,s){\rm e}^{-\frac{|x|}{\sqrt{\tilde{D}_\alpha}}s^{\alpha/2}}\right]\nonumber\;.
\end{eqnarray}
Calculating $\hat{P}(0,s)$ from (\ref{a4}), we finally obtain Eq. (\ref{eq36}).

\end{document}